\documentclass[12pt,notitlepage]{article}

\usepackage{amsmath}
\usepackage{amssymb}
\usepackage{bm}
\usepackage{graphicx}
\usepackage{amsbsy}

\begin{document}


\begin{flushleft}
{\Large
\textbf\newline{A universal rank-size law} 
}
\newline
\\
Marcel Ausloos \textsuperscript{1,2,$\natural$}, Roy Cerqueti
\textsuperscript{3,$\natural$,*}
\\
\bigskip
\textbf{1} School of Business, University of Leicester, University
Road. Leicester, LE1 7RH, United Kingdom. Email: ma683@le.ac.uk\\
\textbf{2}  GRAPES -- Group of Researchers for Applications of
Physics in Economy and Sociology. Rue de la Belle Jardini\`ere 483,
B-4031, Angleur, Belgium. Email: marcel.ausloos@ulg.ac.be
\\
\textbf{3} University of Macerata, Department of Economics and Law.
Via Crescimbeni 20, I-62100, Macerata, Italy. Tel.: +39 0733 258
3246; Fax: +39 0733 258 3205. Email: roy.cerqueti@unimc.it.
\\
\bigskip
$\natural$ These authors contributed equally to this work.

* Corresponding author

\end{flushleft}
\section*{Abstract}

A mere hyperbolic law, like the Zipf's law power function, is often
inadequate to describe rank-size relationships. An alternative
theoretical distribution is proposed based on theoretical physics
arguments starting from the Yule-Simon distribution. A modeling is
proposed leading to a universal form. A theoretical suggestion for
the "best  (or optimal) distribution", is provided  through an
entropy argument.    The ranking of areas through the number of
cities   in various countries and some sport competition ranking
serves for the present  illustrations.


\section{Introduction}
Approaches of hierarchical type lie behind the extensive use of
models in theoretical physics \cite{Hansen14}, the more so when
extending them into new "applications" of statistical physics ideas
\cite{roehner2007driving,stauffer04}, e.g. in complex systems
\cite{kwapien2012physical} and phenomena, like in fluid mechanics
\cite{nonl4,gadomski2000kinetic} mimicking agent diffusion.  In
several studies, researchers have detected the validity of power
laws, for a number of characteristic quantities of complex systems
\cite{West,z1,pnas1,pnas2,ausloos2010physa}. Such studies, at the
frontier of a wide set of scientific contexts,  are sometimes tied
to several issues of technical nature or rely only on the
exploration of distribution functions. To go deeper is a fact of
paramount relevance, along with the exploration of more grounding
concepts.

The literature dealing with the rank-size rule is rather wide:
basically, papers in this field discuss why such a rule should work
(or does not work). Under this perspective, Pareto distribution and
power law, whose statement is that there exists a link of hyperbolic
type between rank and size, seem to be suitable for this purpose. In
particular, the so-called first Zipf's law \cite{z1}, which is the
one associated to a unitary exponent of the power law, has a
relevant informative content, since the exponent can be viewed as a
proxy of the balance between outflow and inflow of agents.

The theoretical explanation of the Zipf's law has been the focus of
a large number of important contributions
\cite{Simon55,gabaix,Gabaix99b,GabaixIoannides04,Brakmanetal99,HillJASA69.74.1017}.
However, the reason   why Zipf's law is found to be a valid tool
for describing rank-sizes rule is still a puzzle. In this respect,
it seems that no theoretical ground is associated to such a
statistical property of some sets of data   \cite{Fujitaetal99,FujitaThisse00}. 

Generally, Zipf's law cannot be
viewed as a universal law, and several circumstances rely on data
whose rank and size relationship is not of hyperbolic nature. Such a
statement is true even in the urban geography case,  - the one of the
original application of the Zipf's law, for the peculiar case of
cities ranking. Remarkable breakdowns has been assessed e.g. in
\cite{CEU31.07.648benguigui,Peng,Matlabaetal, ZDMA,jsmte,pnas3}. A
further example is given by the number $N_{c,p}$ of cities ($c$) per
provinces ($p$) in Italy ($\sum_{c,p} N_{c,p}=$ 8092), see the
log-log plot of the data from 2011 in Fig.
 \ref{fig:Plotnogood9ZMPpwldexp}: the (110) provinces are ranked by
decreasing order of "importance", i.e. their number of cities. Fits
by  (i) a power law, (ii) an exponential and (iii) a Zipf-Mandelbrot
(ZM)   function \cite{FAIRTHORNE}

\begin{equation} \label{ZMeq3}
y(r)=\hat{c}/(\rho+r)^{\omega} \;\equiv \; [c/(\rho+r)]^{\omega},
\end{equation}
$r$  being the rank.

The fits  are, the least to say, quite unsatisfactory in particular
in the high rank tail, essentially because data usually often
presents  an inflection point.

Therefore, no need to elaborate  further that more data analysis can
bring some information on   the matter.

The paper is organized as follows.  In Section \ref{alternative}, an
alternative to a hyperbolic rank-size law  and its above
"improvements" are  discussed:   the data  can be better represented
by an  (other than  Zipf's law)  analytic   empirical law,
 allowing for an inflection point.

Next, we  introduce a universal form,  allowing for a wider appeal,
in Sect. \ref{universal},  based on a  model  thereafter presented
in Sect. \ref{modelBeta}. Such a general law can be turned into a
frequency or probability distribution. Thus, the method suggests to
consider  a criterion of possibly optimal organization  through the
notion of relative distance to full disorder, i.e., a  ranking
criterion of entity distributions based on the entropy (Section
\ref{sec:entropy}). Section \ref{conclusions} allows us to conclude
and to offer suggestions for further research lines.

\section{An alternative to a hyperbolic rank-size law }\label{alternative}

\begin{figure}[!h] 
\includegraphics[height=15.8cm,width=12.2cm]{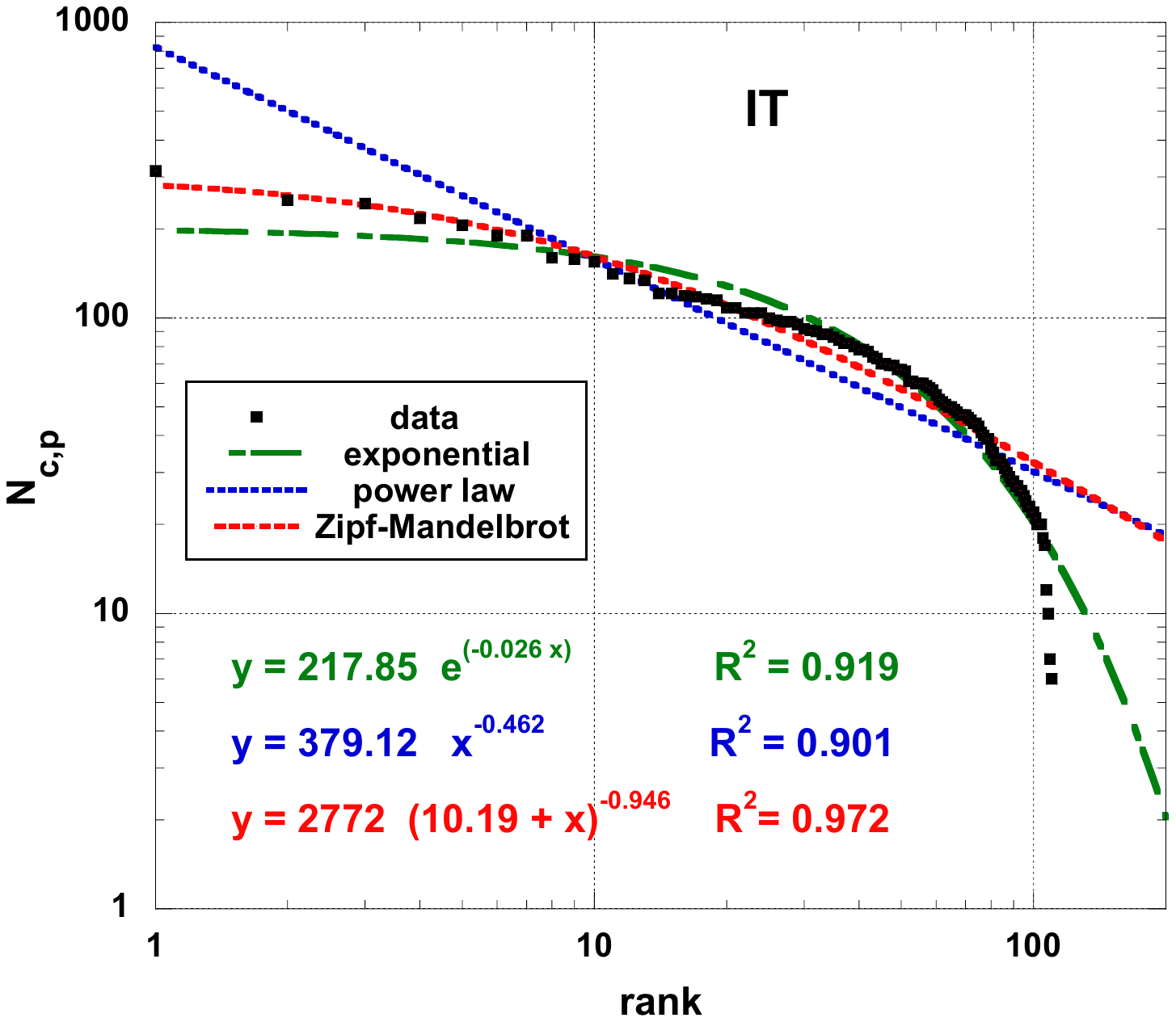}
\caption{Relationship between the number $N_{c,p}$ of IT cities
(8092) per  provinces (110) on a log-log scale. The ranking
criterion is the one associated to the number of cities (high rank
when the number of cities is high). The reference year is 2011.
Several fits are shown: power law, exponential and Zipf-Mandelbrot
function, Eq. (\ref{ZMeq3}). The corresponding correlation
coefficients are given; different colors and symbols allow to
distinguish cases.} \label{fig:Plotnogood9ZMPpwldexp}
\end{figure}

In the context of best-fit procedures, rank-size theory allows to
explore the presence of regularities among data and their specified
criterion-based ranking \cite{Jefferson39}. Such regularities are
captured by a best-fit curve. However, as observed in Fig.
 \ref{fig:Plotnogood9ZMPpwldexp}, the main problem strangely resides
in missing the distribution high rank tail behavior. No doubt, that
this partially arises because most fit algorithms take better care
of the high values (on the $y$-axis) than the small ones.  More
drastically, a cause stems in the  large rank $r$ tail which is
usually supposed to extend to infinity, see Eq.
(\ref{PWLwithcutoff}), but  each system is markedly always of finite
size. Therefore,  more complicated laws containing a power factor,
like  the stretched exponential or exponential cut-off  laws should
be considered inadequate.

We emphasize that we are in presence of  data which often exhibits an
inflection point.

The presence of an inflection point means that there is a change in
the concavity of the curve, even if the slope remains with the same
(negative) sign for the whole range. Thus, one could identify two
regimes in the ranked data, meaning that the values are clustered in
two families at a low and high ranks. In such cases, the finite
cardinality $N$ of the dataset leads to a collapse of the upper
regime at rank $r_M\equiv N$.  Nevertheless, the Yule-Simon
distribution \cite{Pwco3},
 \begin{equation} \label{PWLwithcutoff}
 y(r)= d \;r^{-\alpha} \; e^{-\lambda r},
\end{equation}
 could be arranged in an
appropriate way, according to a Taylor expansion as in \cite{jsmte}.
Eq. (\ref{PWLwithcutoff}) can be then rewritten as
\begin{equation} \label{Lavalette3a}
y(r)= \kappa_3\;  \frac{(N\;r)^{- \gamma}}  { (N-r+1)^{-\xi}  },
\end{equation}
as discussed by Martinez-Mekler et al. \cite{ Mekler} for
rank-ordering distributions, - in the arts and sciences; see also
more recent work on the subject  \cite{1606.01959v1LavalbeyondZipfMiramontes} with references therein.
 This Eq. (\ref{Lavalette3a}) is a (three-parameter) generalization of
  \begin{equation} \label{Lavalette2}
 y(r)=K\;  \Big(\frac{N\cdot r}{N-r +1}\Big)^{-\beta}\;\equiv\; \kappa_2\;  \Big(\frac{r }{N-r +1}\Big)^{-\beta},
 \end{equation}
the (t-parameter) function used when considering the distribution of
impact factors in bibliometrics studies
\cite{RRPh49.97.3popescu,Glottom6.03.83popescu,JoI1.07.155Mansilla,JQL18.11.274Voloshynovska,MAJAQM},
i.e., when $   \gamma \equiv \xi$,  and recently applied to
religious movement adhesion \cite{MALavPRE}.   Notice that there is
no fundamental reason why the  decaying behavior at low rank should
have the same exponent as the collapsing regime at high rank:  one
should  {\it a priori} admit  $\gamma$ $\neq$  $\xi$.

\begin{figure}[!h]
 \includegraphics  [height=15.8cm,width=12.2cm]{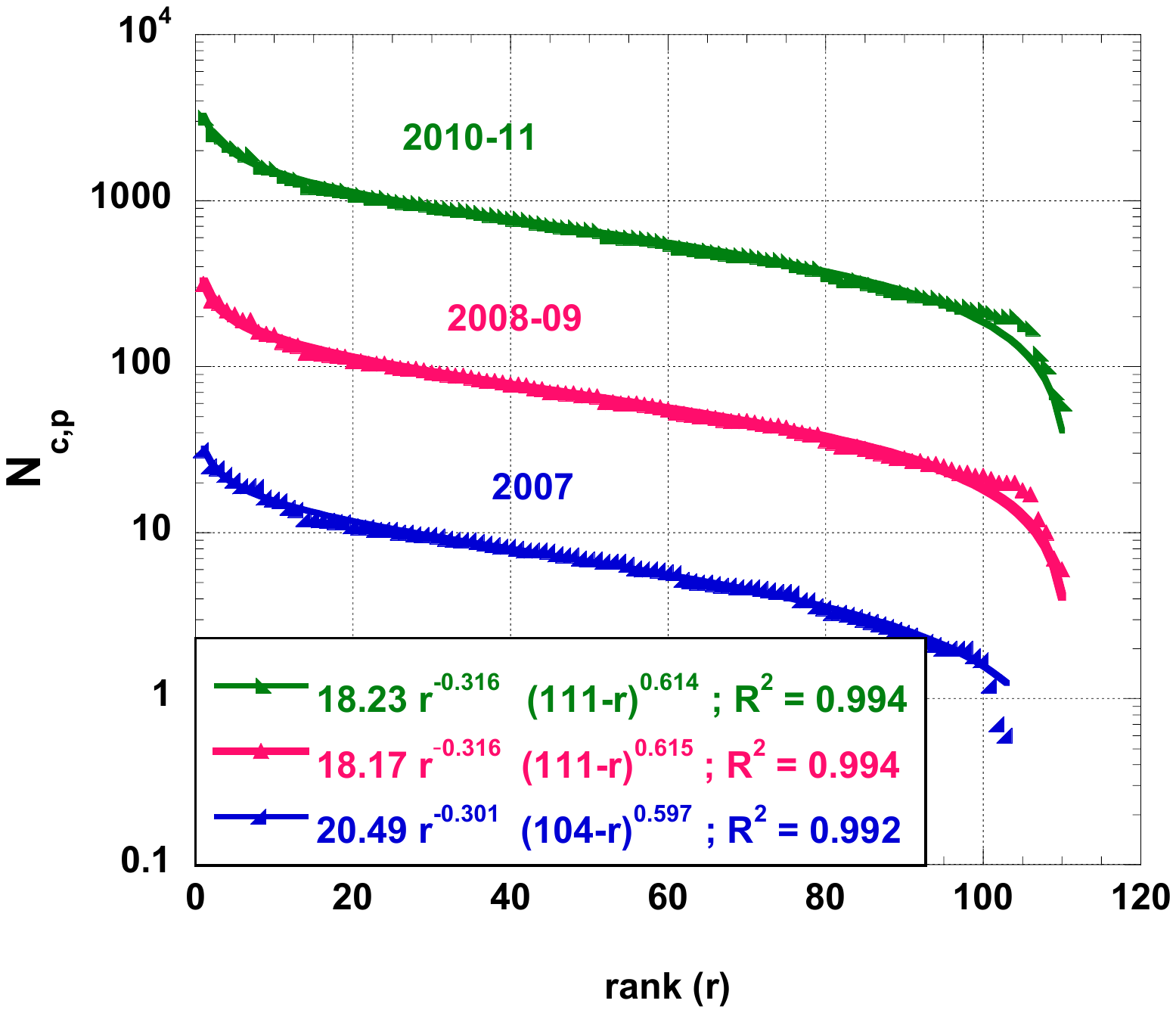}
 \caption   {Semi-log  plot of the number of   cities  in IT provinces, $N_{c,p}$; the provinces are  ranked by their
decreasing "order of  importance", for various years;   the  2007,
2008-2009 and 2010-2011;  data are displaced by an obvious factor
for better readability; the best  3-parameter   function, Eq.
(\ref{Lavalette3a}),  fit is   shown. Parameter values are obtained
by fits through Levenberg-Marquardt algorithms with a 0.01\%
precision.}
 \label{fig:PlotliloNcp3f}
\end{figure}

In fact, such an alternative law is   easily demonstrated to be an
appropriate one for describing size-rank data plots. For example,
reconsider  the IT  $N_{c,p}$  case,  shown in Fig.
 \ref{fig:Plotnogood9ZMPpwldexp},  redrawn on a semi-log plot in Fig.
 \ref{fig:PlotliloNcp3f}. (All fits, in this communication, are based on the Levenberg-Marquardt algorithm
\cite{Levenberg1944,Marquardt1963,Lourakis2011} with a 0.01\%
imposed precision and after testing various initial conditions for
the regression process.)  The rank-size relationship appears to
follow a flipped noid function around some horizontal mirror or
axis. Notice that  
similar behaviors are observed for different years, although  the
number of $N_{c,p}$ yearly differs.  Incidently,  note that, in this
recent time, the  official data claims a number of 103 provinces in
2007, with an increase by 7 units (BT, CI, FM, MB, OG, OT, VS, in
conventional notations) thereafter, leading to 110 provinces. The
number of municipalities has also been changing, between 2009 and
2010, whence the rank of a given province is not constant over the
studied years.
\subsection{Other illustrating  topics and generalization }\label{universal2}

In view of taking into account a better fit at low and high rank,
one can generalize Eq. (\ref{Lavalette3a}) to a five parameter free
equation

\begin{equation} \label{Lavalette5}
y(r)= \kappa_5\;  \frac{(N\;(r+\Phi))^{- \gamma}}  { (N+1-r+\Psi)^{-\xi}  },
\end{equation}
where the parameter $\Phi$ takes into account Mandelbrot
generalization of Zipf's law at low rank,  see Eq. (\ref{ZMeq3}),
while $\Psi$ allows some flexibility at the highest rank.

In particular, the shape of the curve in Eq. (\ref{Lavalette5}) is
very sensitive to the variations of $\Phi$ and $\Psi$.

As the parameter $\Phi$ increases, the relative level of the sizes
at high ranks is  also increased. This means that the presence of
outliers at high ranks is associated to high  values of $\Phi$. If
one removes such outliers from the dataset and implements a new fit
procedure,
 one obtains a lower level of the calibrated $\Phi$ and a
flattening of the curve at  low ranks. In
\cite{AusloosandCerqueti15}, the authors have found something
similar in a different context;  they have denoted the major
(upsurging) outlier at rank $1$ by "king" and called  the other
outliers at ranks $2,3,\dots$ as "viceroys". The removal of outliers
necessarily leads to a more appealing  fit , in terms of
visualization and $R^2$, when such a procedure is implemented
through
 power laws. In this respect, the introduction of a further
parameter -- $\Phi$, in this case -- serves as adjustment term at
high ranks, and represents an improvement of the previous theory.

Indeed, the parameter $\Psi$ acts analogously  to  $\Phi$, but at a
low rank. In particular, an increase of $\Psi$ is associated to a
flattening of the five parameter curve of Eq. (\ref{Lavalette5}) at
medium and low   ranks. Such a flattening is due to sizes at low
ranks which are rather close to those at  medium ranks. This
phenomenon has been denoted  in \cite{Ausloos13} as "queen" and
"harem" effect, - to have in mind the corresponding "king" and
"viceroys" effects at  low ranks. The queen and harem effect is
responsible of the deviations of the power law from the empirical
data at a low rank. Thus, the parameter $\Psi$ also constitutes an
adjustment term at low ranks and is an effective improvement of the
performance of the fitting procedure.

Substantially, the specific sense of $\Psi$ should be also  read in
terms of "generalization"  and "in view of best fit". Usually, one
is not sure about the 0 at the origin of axes. Our $\Phi$
corresponds to the $\rho$ of Mandelbrot (see Eq. (\ref{ZMeq3})), for
which Mandelbrot gives no interpretation: it is only a mathematical
trick. Thus, by "symmetry", we introduce  a $\Psi$ at high rank. It
allows some flexibility due to possible sharp decays, due to
outliers at high ranks. This also allows to move away from strict
integers, and open the functions to continuous space as done in
Sect. \ref{universal}.

\begin{figure}[!h]
 \includegraphics  [height=15.8cm,width=12.2cm]{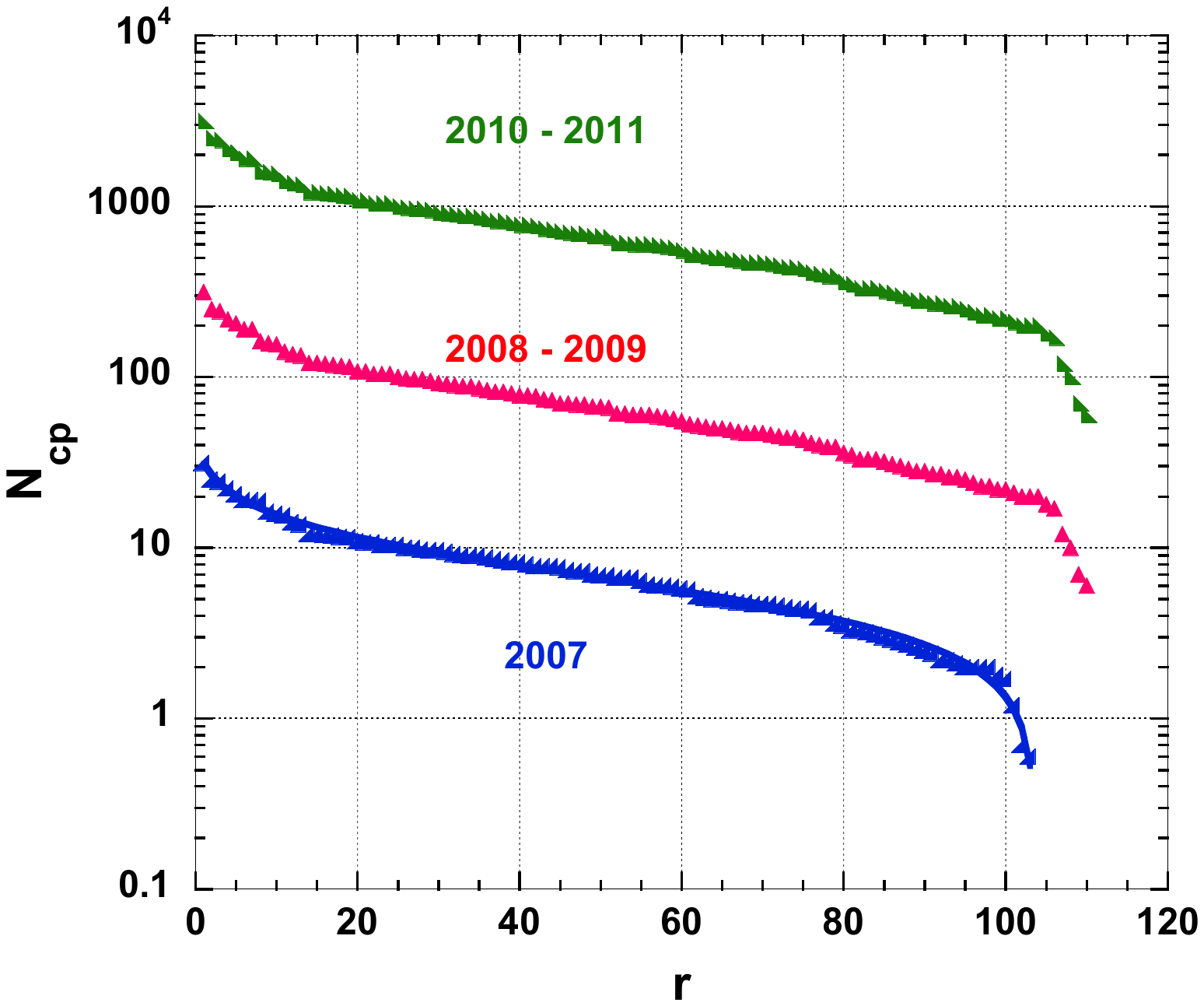}
 \caption   {Semi-log  plot of the number of   cities in IT provinces, $N_{c,p}$; the provinces are ranked
 according to their
decreasing "order of  importance", for various years;   the  2007
 and 2010-2011 data are displaced by an obvious factor of 10
for better readability; the best  5-parameter   function, Eq.
(\ref{Lavalette5}),  fit is   shown. Parameter values are obtained
by fits through Levenberg-Marquardt algorithms with a 0.01\%
precision.}
 \label{2nd-5}
\end{figure}

We have compared the fits conceptualized in Eq. (\ref{Lavalette3a})
and Eq. (\ref{Lavalette5}) for the specific IT $N_{c,p}$ case
(compare Fig.  \ref{fig:PlotliloNcp3f} and Fig. \ref{2nd-5} and
Table \ref{table2nd}). Even if both these laws are visually
appealing and exhibit a high level of goodness of fit, the $R^2$
associated to Eq. (\ref{Lavalette3a}) is slightly lower than that of
Eq. (\ref{Lavalette5}). Thus, we can conclude that the
five-parameters law, Eq. (\ref{Lavalette5}),  performs better than
the three-parameters one, Eq. (\ref{Lavalette3a}).

\begin{table}[!ht]
\centering \caption{Best fit parameters and $R^2$ for
 the number of IT cities in  various years, for either  cases, i.e. Eqs.
(\ref{Lavalette3a}) and (\ref{Lavalette5}).  The parameters pertain
to the displaced data as visualized on
 Fig. \ref{fig:PlotliloNcp3f} and  Fig. \ref{2nd-5}.}
\begin{tabular}{|l|l|l|l|}
   \hline
\multicolumn{4}{|c|}{\textbf Eq. (\ref{Lavalette3a})}\\
\hline
  &2007 &2008/2009 & 2010/2011   \\
\hline
 $N$&103&110&110 \\
 $\kappa_3 N^{-\gamma}$&2.049&18.177&182.265 \\
 $\gamma$&0.301&0.316&0.316  \\
 $\xi$&0.597&0.615&0.614\\ \hline
$R^2$  &0.99240& 0.99445& 0.99441    \\ 
\hline
\multicolumn{4}{|c|}{\textbf Eq. (\ref{Lavalette5})}\\
\hline
&2007 &2008/2009 & 2010/2011   \\
\hline
$N$&110&110&110 \\
  $ \kappa_5 N^{-\gamma} $&3.971&33.709&332.71 \\
 $\gamma$&0.373&0.387&0.386  \\
 $\xi$&0.499&0.527&0.529\\
  $\Psi$&-7.441&0.608&0.640\\
  $\Phi$&0.945&0.926&0.906\\
   \hline
$R^2$  &0.99402& 0.99631& 0.99623 \\ \hline \end{tabular}
\begin{flushleft}
\end{flushleft} \label{table2nd} 
\end{table}

Even though one could display many figures describing the usefulness
of the  above, let us consider two cases, e.g. 
 in  sport matter.

\begin{figure}[!h]
 \includegraphics[height=15.8cm,width=12.2cm]{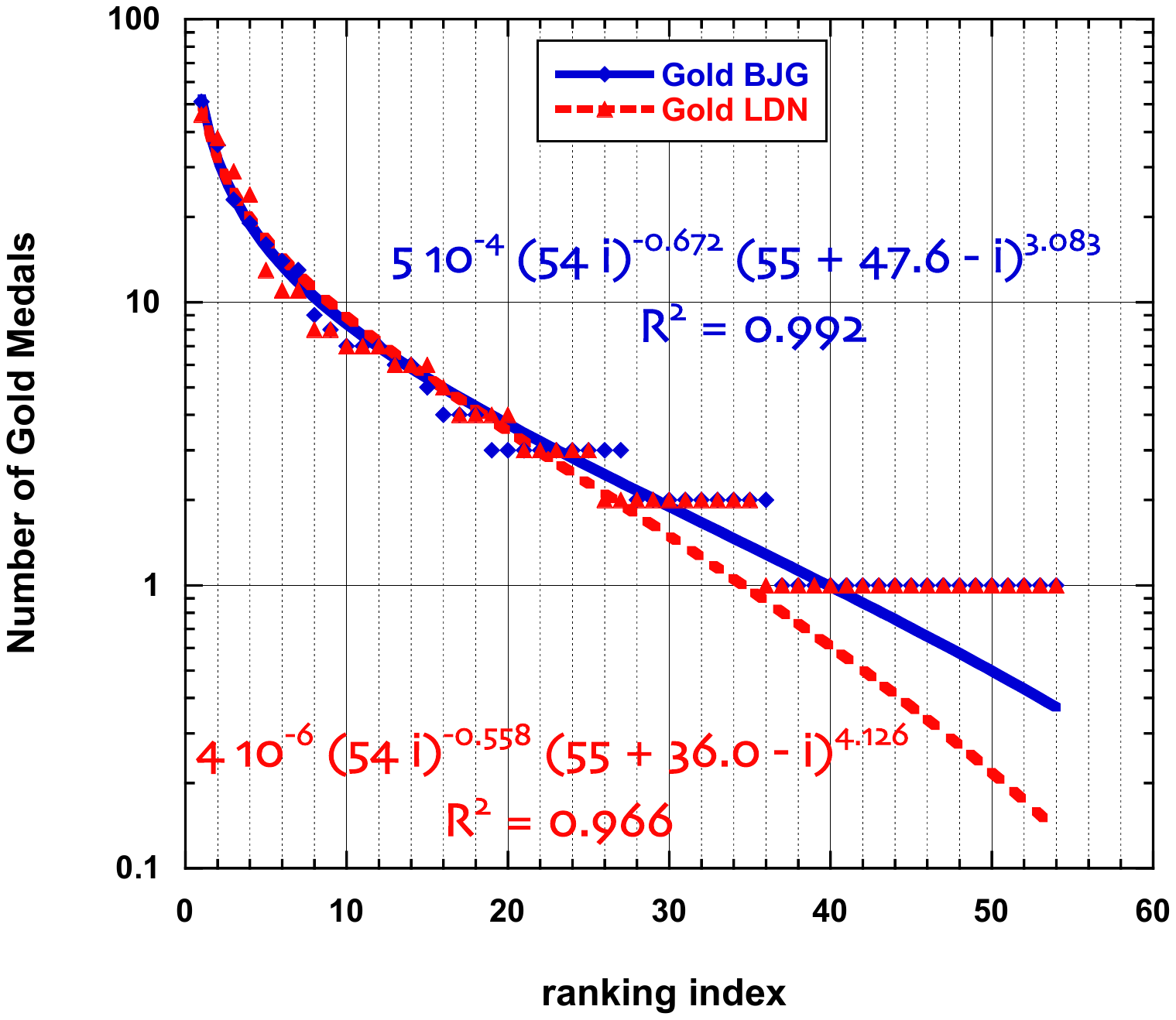}
 \caption   {Semi-log  plot of the number of Gold medals obtained by countries  at Beijing (BJG) and London (LDN) recent Summer Olympic Games, as ranked according to their
decreasing "order of  importance" index $i$;  the best   4-parameter
fitting  function  is displayed, Eq. (\ref{Lavalette5}),  with
$\Phi=0$. }
 \label{fig:Plot15GoldBjGLDNLav4}
\end{figure}

\begin{figure}[!h]
 \includegraphics[height=15.8cm,width=12.2cm]{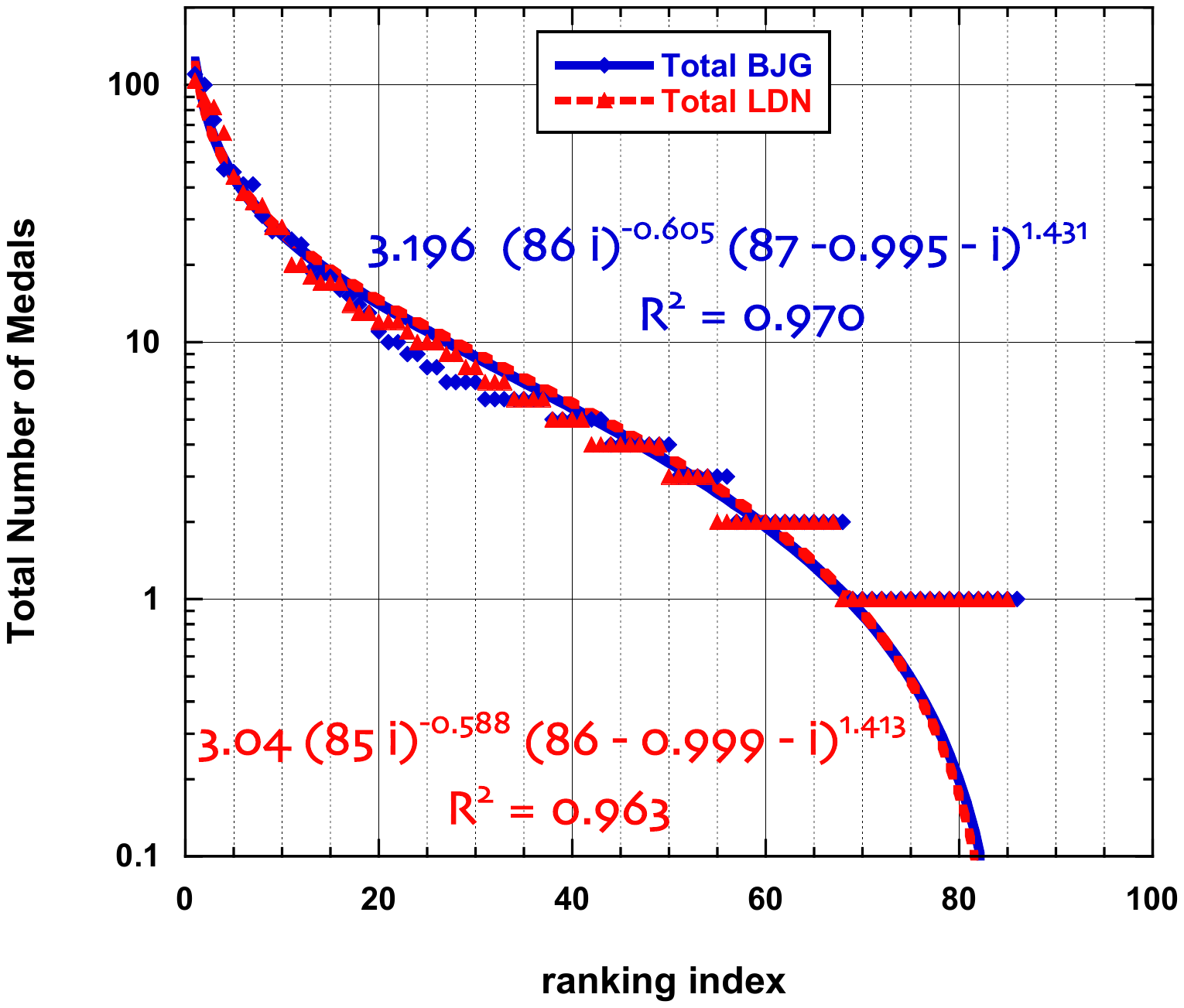}
 \caption   {Semi-log  plot of the total number of   medals  obtained by countries  at Beijing (BJG) and London (LDN) recent Summer Olympic Games, as ranked according to their
decreasing "order of  importance" index $i$;  the best   4-parameter
fitting  function  is displayed, Eq. (\ref{Lavalette5}),  with
$\Phi=0$. }
 \label{fig:Plot15TotalBjGLDNLav4}
\end{figure}

\begin{figure}[!h]
 \includegraphics   [height=15.8cm,width=12.2cm]{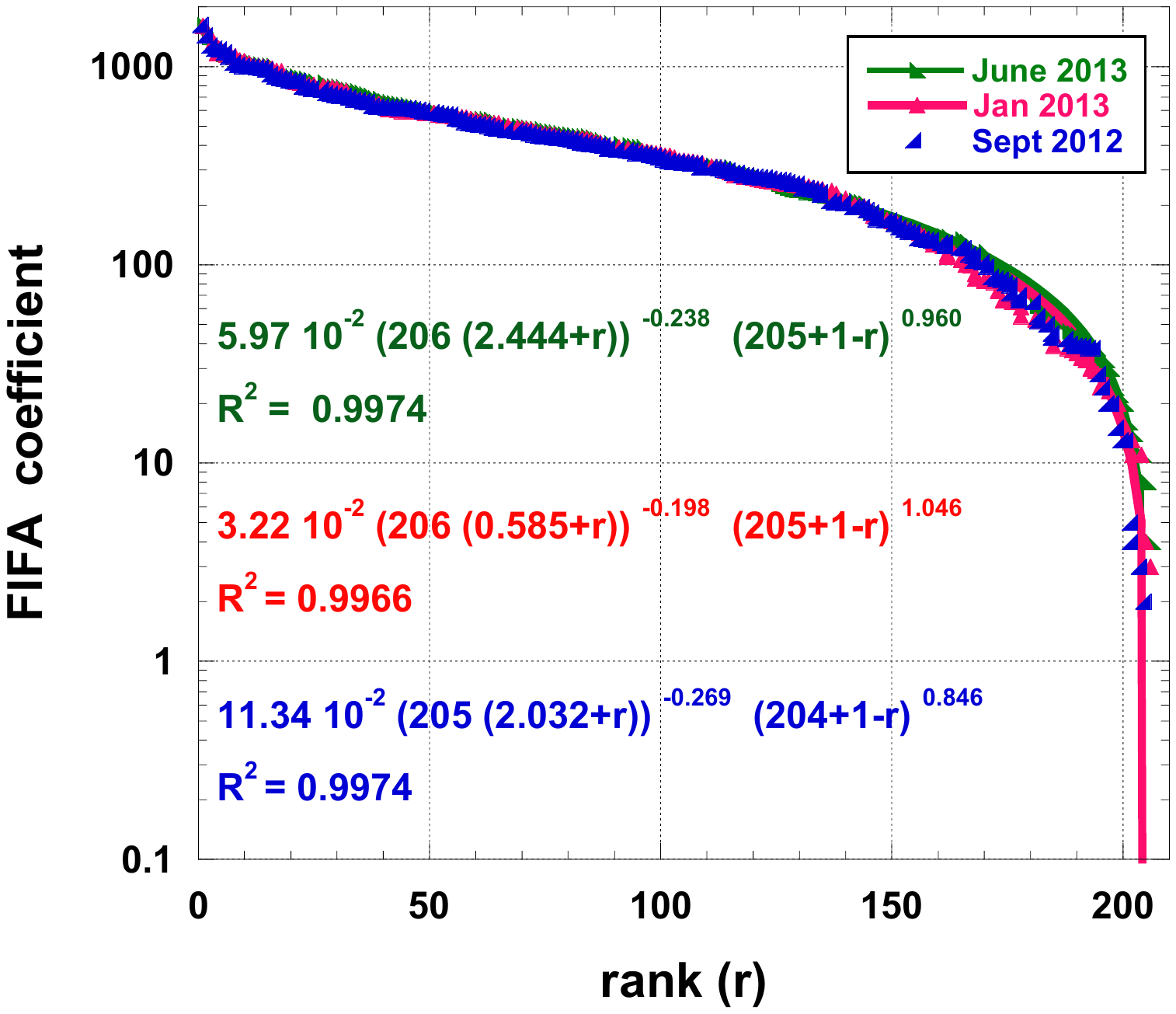}
 \caption   {Semi-log  plot of the FIFA countries ranked by their
decreasing "order of  importance" through the FIFA coefficient;  the
best   5-parameter   function, Eq. (\ref{Lavalette5}),  is   shown.
}
 \label{fig:Plot61FIFA1213Lav5liloN206}
\end{figure}

\begin{itemize}\item
Consider the ranking of countries at recent Summer Olympic  Games:
Beijing  2008 and London 2012. The ranking of countries is performed
trough the number of "gold medals", but one can also consider the
total number of medals, - thus considering a larger set of
countries.  A country rank  is of course varying according to the
chosen criterion.  It is also true that due to subsequent analysis
of athlete urine and other doping search tests, the attribution of
medals may change with time. We downloaded  the data available on
Aug. 13, 2012,   from

  $ http://www.bbc.co.uk/sport/olympics/2012/medals/countries$.

Interestingly, the number of gold medals has not changed between
Beijing and London, i.e. 302, but due to the "equivalence of
athletic scores", the total number of medals is slightly different :
958 $ \rightarrow$ 962. Moreover, the number of countries having
received at least a gold medal is the same (54), but the total
number of honored countries  decreased from 86 to 85. Obviously, in
contrast to the administrative data on IT provinces ranking, there
is much "equality between countries" in Olympic  Games; therefore a
strict rank  set contains many empty subsets.  It is common to
redefine a continuous  (discrete) index $i$ in order to rank the
countries. Moreover, the rank  distributions are much positively
skewed (skewness $\sim$ 3)  with high kurtosis ($\ge 10$).
Therefore, the inflection points occur near $r=r_M/2$ and for  a
size  close to the median value. On Fig.
 \ref{fig:Plot15GoldBjGLDNLav4}  and
Fig. \ref{fig:Plot15TotalBjGLDNLav4}, such a ranking for Olympic
Games medals is   displayed, both for the Gold medal ranking and the
overall  ("total") medal ranking. Reasonably imposing $\Phi=0$, the
parameters of Eq. (\ref{Lavalette5})   lead to remarkable fits, even
though the collapsing behavior of the function occurs outside the
finite $N$ range. We have tested that  a finite $\Phi $ does not
lead to much  regression coefficient  $R^2$ improvement.

\item In other sport competitions, the "quality" of teams or/and  countries  is measured through quantities which are not discrete values.
For example, in soccer,  more than 200  federations (called
"Association Members", $\sim$ countries) are affiliated to the FIFA

($http://www.fifa.com/worldranking/procedureandschedule/menprocedure/index.html$).
The FIFA Country ranking system  is based on results over the
previous four years since July 2006. It is described and discussed
in \cite{IJMPCFIFAMARCAGNV} to which we refer the reader for more
information. Note that a few countries have zero FIFA coefficients.
Interestingly the skewness and kurtosis of the FIFA coefficient
distributions are rather "well behaved" (close to or $\le$1.0),
while the  coefficient of dispersion is about  250. From previous
studies, it can be observed that the low rank ("best countries")
are well described by a mere power law, including the Mandelbrot
correction to the Zipf's law. However, the high tail behavior is
poorly described. We show in Fig.
\ref{fig:Plot61FIFA1213Lav5liloN206} that the generalized equation
is much better indeed. From a sport analysis point of view, one
might wonder about some deviation in the ranking between 170 and
190.

\end{itemize}

\subsection{Universal form }\label{universal}
These displays  suggest  to  propose some universal vision as presented
next.

It is easily observed  in Eq. (\ref{Lavalette3a})
 that a change of variables
 $u \; \equiv \;  r/(N+1)$,
 leads to

\begin{equation} \label{Lavalette3u}
y_1(u)= \hat{\kappa_3}\;   \Big[ u^{-\gamma} (1-u)^{\xi}  \Big]
\end{equation}
 
However, in so doing, $u\in [ 1/(N+1), N/(N+1)]$.

In order  to span the full $[0,1]$ interval, it is better to
introduce the reduced variable   $w$,  defined as $w\equiv
(r-1)/(r_M-1)$,  where $r_M$ is the maximum number of entities.
Moreover, in order to fully generalize the empirical law, in the
spirit of ZM, Eq. (\ref{ZMeq3}),  at low rank,  a parameter $\phi$
can be introduced. In the same spirit, we admit a fit parameter
$\psi$ allowing for   possibly   better convergence at $u\simeq1$;
we expect, $\mu \sim 1/r_M$.

Thus, we propose the universal form
\begin{equation} \label{Lavalette5u}
y_2(w)=  \eta\;\; (\phi+w)^{-\zeta}\;\;\Big[1-w+\psi\Big]^{\chi},
\end{equation}
for which the two exponents $\chi$ and $\zeta$ are the theoretically
meaningful parameters. The amplitude $\eta$ represents a normalizing
factor, and can be then estimated. Indeed, by referring to the case
$\chi \in (0,+\infty)$ and $\zeta\in (0,1)$ and posing
$\tilde{w}=\phi+w$ and $u=1+\phi+\psi$, we can write
  \begin{eqnarray} \label{etaLavalette5ri}\nonumber
\eta=\Big[  \int_{\tilde{w}_0}^{\tilde{w}_1}  \tilde{w}^{-\zeta}
(u-\tilde{w})^{\chi} \;d\tilde{w} \Big]^{-1} \equiv \\
\frac{1}{(1+\phi+\psi)^{1+\chi-\zeta}}\;\;\frac{1}{[B_{t}(1-\zeta,1+\chi)]_{t_0}^{t_1}}
\;\;,
\end{eqnarray}
with $t_0 = \phi/(1+\psi+\phi)$, and  $t_1 =(1+
\phi)/(1+\psi+\phi)$, and where $B_t (x, y) $ is the incomplete
Euler Beta function \cite{AbraSteg,GradRyz,PearsonBTables}, itself
easily written, when $t=1$,  in terms of the Euler Beta function,
\begin{equation}
B(x, y)\equiv B_1(x, y) = \frac{ \Gamma(x) \Gamma(y)}{\Gamma(x+y)};
\end{equation}
$\Gamma(x)$ being the standard Gamma function.

\begin{figure}[!h]
 \includegraphics[height=15.8cm,width=12.2cm]{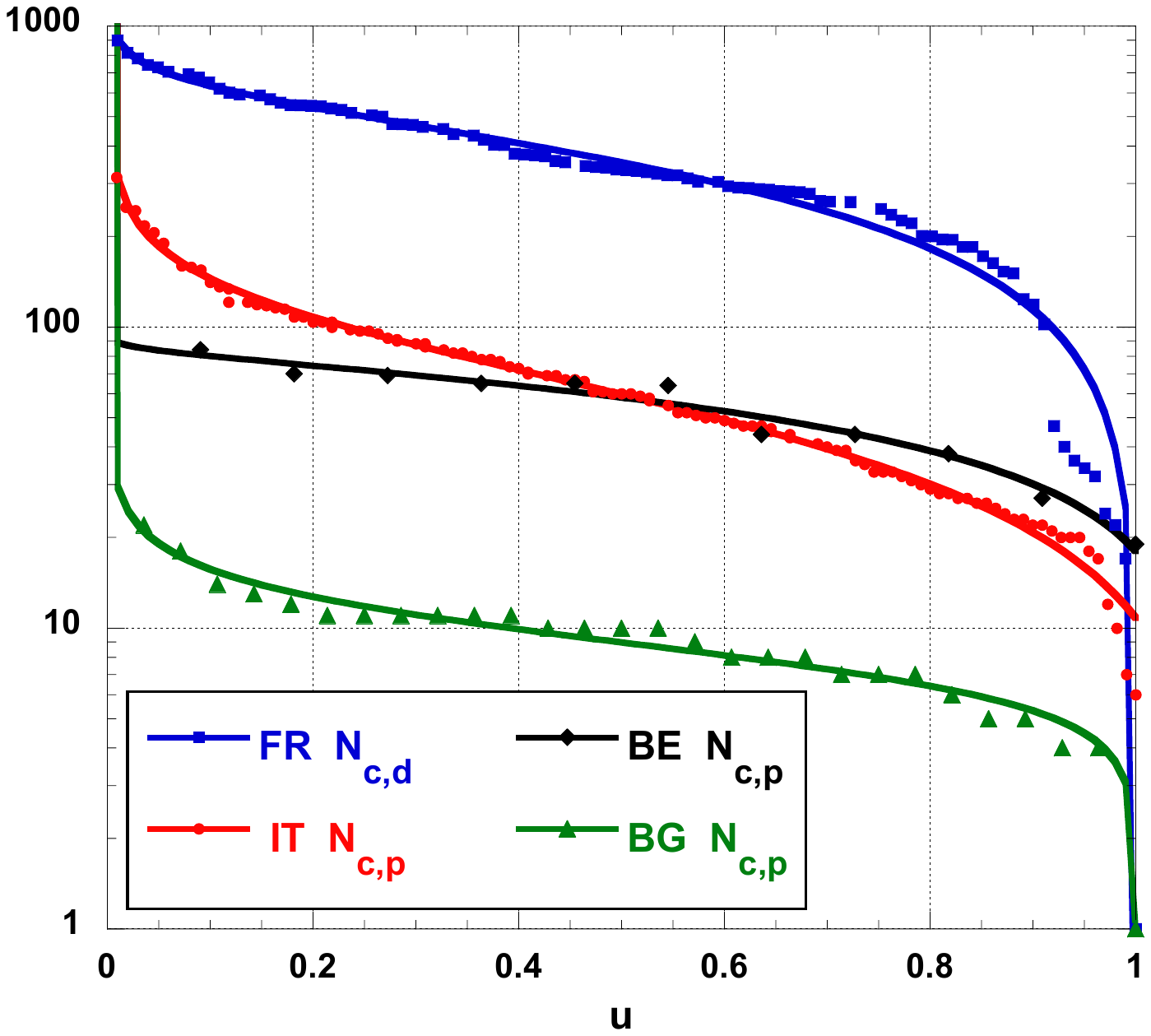}
 \caption   {Semi-log plot of the number  of cities, $N_{c,p}$  and $N_{c,d}$,
ranked by decreasing order of "importance" -in the sense of "number
of cities"- of provinces (in BE, BG, and IT) or departments (in FR);
the best function fit, Eq. (\ref{Lavalette5u}), is   shown;
parameter values are found in Table \ref{Tablecitprovreg}.}
 \label{fig:Plot23BEGFRITuNcpdL}
\end{figure}

The function in Eq. (\ref{Lavalette5u})  is shown  on Fig.
\ref{fig:Plot23BEGFRITuNcpdL} to describe different cases, with
various orders of magnitude,  i.e.,   a semi-log plot of the number
of cities  in a province, $N_{c,p}$ or in a  department, $N_{c,d}$,
ranked by decreasing order of "importance", for various countries
(BE, BG, FR, IT). The reference year is 2011. In such cases,
$\phi\equiv0$, obviously, thereby much simplifying Eq.
(\ref{etaLavalette5ri}), whence reducing the fit  to a three free
parameter search.

For completeness,  the main statistical indicators for the number of
cities  ($N_c $),   in the  provinces ($N_{c,p}$), regions
($N_{c,r}$) or departments  ($N_{c,d}$)  in these (European)
countries, in 2011  is given in   Table \ref{Tablecitprovreg}.
Notice that the distributions differ:  the median  ($m$) is
sometimes larger (or smaller) than the mean ($\mu$), while the
kurtosis and skewness can be positive or negative. Yet the fits with
Eq. (\ref{Lavalette5u})  seem very fine. The large variety in these
characteristics  is an {\it a posteriori} argument in favor of
having examined so  many cases.

\subsection{Modelization}\label{modelBeta}

The presented argument  is of wide  application as the reader can
appreciate. However, the vocabulary in this modeling section can be
adequately taken from  the jargon of city evolution for better
phrasing and for continuing with the analyzed data.

A  preferential attachment process can be defined as a settlement
procedure in urn theory, where additional balls are added and
distributed continuously to the urns (areas, in this model)
composing the system. The rule of such an addition follows an
increasing function of the number of the balls already contained in
the urns.

In general, such a process contemplates also the creation of new
urns. In such a general framework, this model is associated to the
Yule-Simon distribution, whose density function $f$ is
 \begin{equation} f(a;b) = b\,B(a, b+1),\end{equation}
being $a$ and $b$ real nonnegative numbers.

The integral $\int^1_0 x^a\;(1-x)^b\;dx$ represents the probability
of selecting $a+b+1$ real numbers such that the first one coincides
with $x$, from the second to the $a+1$-th one numbers are less or
equal to $x$ and the remaining $b$ numbers belong to $[x,1]$.

In practical words, newly created urn starts out with $k_0$ balls
and further balls are added to urns at a rate proportional to the
number $k$ that they already have plus a constant $a\ge -k_0$.  With
these definitions,   the fraction $P(k)$ of urns (areas)  having $k$
balls (cities)  in the limit of long time is given by
\begin{equation}
 P(k) = \frac{ B(k+a;b)}{B(k_0+a;b-1)} \end{equation} for $k\ge0$ (and zero otherwise).
In such a limit, the preferential attachment process generates a
"long-tailed" distribution following  a  hyperbolic (Pareto)
distribution, i.e.  power law, in  its tail.

It is important to note that the hypothesis of continuously
increasing urns is purely speculative, even if it is widely adopted
in statistical physics. Indeed, such an assumption contrasts with
the availability of resources, and the growth of the number of
settlements is then bounded. Therefore, as in Verhulst's
modification \cite{Vlog3} of the Keynesian expansion model of
population, a  "capacity factor" must be introduced in  the original
Yule process, thereby leading to the $u$ term in Eq.
 (\ref{Lavalette5}) and its subsequent interpretation.
 
\section*{Entropy connection}\label{sec:entropy}

One can consider to have  access to a sort of  "probability" for
finding a certain "state"  (size occurrence)  at a certain rank,
through
 \begin{equation} \label{pr}\nonumber
p(w)  \sim  y_2(w)\sim \frac{(\phi+w)^{-\zeta}(1-w+\psi)^{\chi}}{
(1+\phi+\psi)^{\chi-\zeta+1}B(\chi+1,1-\zeta)}\;,
\end{equation}
the denominator  resulting from Eq. (\ref{etaLavalette5ri}).

Thereafter, one can obtain something which looks like
the Shannon entropy \cite{shannon} : $ S\equiv   -\int p(w)\; ln
(p(w))$. It has to be compared to the maximum disorder number,  i.e.
$ln (N)$. Whence we define the relative distance  to the maximum
entropy as  \begin{equation} \label{d}d= \frac {S
}{ln(N)}-1.\end{equation} As a illustration, the only case of the
ranking of cities in various countries is discussed. Values are
reported in Table \ref{Tablecitprovreg}. It is observed that the FR
and IT $d$-values are more extreme than those of BG and BE.  This
corroborates  the common knowledge that the former two countries
have too many cities, in contrast to the latter two.

Thus, in this particular case, this distance concept based on the
universal ranking function with the two exponents $\zeta$ and $\chi$
shows its interest, e.g. within some management  or control process.
It can be conjectured without much debate that this concept can be
applied in many other cases.

It is relevant to note that the entropy argument can be extended in
a natural way to the $q$-Tsallis statistics analysis. Such an
extension could add further elements to the thermodynamic
interpretation of the proposed rank-size analysis. More in details,
rank-size law might be associated to $q$-Tsallis distribution
through a generalization of the central limit theorem for a class of
non independent random variables (see e.g. \cite{moyanotsallis} and
\cite{naumiscocho2007}).
 However, the Tsallis approach is
well-beyond the aim of the present study, and we leave this issue to
future research.

\section{Conclusions}\label{conclusions}
This paper provides a  basically three parameter function  for the
rank-size rule, based on preferential attachment considerations and
strict  input of finite size sampling.  The analysis of the
distribution of municipalities in regions or  departments has proven
the function value  after its mapping into "dimensionless
variables". It seems obvious that the approach is very general and
not limited to this sort of data. Other aspects suggest to work on
theoretical improvements of   the  rank-size  law connections,
through ties with   thermodynamics features, e.g., entropy  and
time-dependent evolution equations ideas.

  \vskip0.2cm {\bf Acknowledgements}
 \vskip0.2cm
 This paper is part of scientific activities in COST Action IS1104,
 "The EU in the new complex geography of economic systems: models,
 tools and policy evaluation" and in COST Action TD1210 'Analyzing
 the dynamics of information and knowledge landscapes'.

\begin{table}[!ht]
\centering \caption{Statistical characteristics of the distribution
of the Number of cities $N_c$,    number of provinces $N_p$   or
departments, $N_d$ (in FR),  in 2011, in 4 European countries;
relevant fit exponents with Eq. (\ref{Lavalette5}), and entropic
distance $d$.} \begin{tabular}{|l|l|l|l|l|l|}
   \hline
   $ $   &BE &BG &FR&  IT&  \\
\hline
$N_c $  &   589 &   264 &   36683   &   8092   &   \\
$N_x$ ($x=p,d$)     &   11  &   28  &   101 &   110 &   \\
Min &   19  &   10  &   1   &   6   &   \\
Max &   84  &   22  &   895 &   315 &   \\
Mean ($\mu$)&   53.55   &   9.429   &   363.2   &   73.56   &   \\
Median  &   64  &   10  &   332 &   60  &   \\
Std Dev ($\sigma$)& 20.32   &   4.273   &   198.3   &   55.34   &   \\
Skewness    &   -0.311  &   0.781   &   0.332   &   1.729   &   \\
Kurtosis    &   -1.045  &   1.480   &   -0.286  &   3.683   &   \\
\hline
$\mu/\sigma$    &   2.635   &   2.207   &   1.832   &   1.329   &   \\
$3(\mu-m)/\sigma$   &   -1.543  &   -0.401  &   0.472   &   0.735 &
\\  \hline
$N$&11&28&101&110&\\
$\kappa_5$&7.49&10.28&84.69&203.8&\\
$\gamma$&-0.160&0.157&0.133&0.386&\\
$\xi$&0.631&0.310&0.653&0.529&\\
$\Psi$&-0.0399&-0.985&-0.999& 0.640&\\
$\Phi$&17.56&-0.820&0.265&0.906&\\
$R^2$  &0.958& 0.975& 0.990 &0.996&    \\\hline
$ln(N_x)$ ($x=p,d$) &      2.3979  &   3.3322  &   4.6151  &   4.7005  &   \\
$d$ &   0.1587 &    0.1959 &    0.3793  &   0.2451  &   \\  \hline
\end{tabular}
\begin{flushleft}
\end{flushleft}
\label{Tablecitprovreg}
\end{table}

\end{document}